\documentclass[sigconf]{acmart}
\usepackage{balance}
\usepackage{helvet}  
\usepackage{courier}  
\usepackage{url}  
\usepackage{graphicx}  
\usepackage{booktabs} 
\usepackage{amssymb}
\usepackage{amsmath}
\usepackage{algorithm}
\usepackage{algorithmic}
\usepackage{amsthm}
\usepackage{multirow}
\usepackage{booktabs}
\usepackage{pgffor}
\usepackage{graphicx}
\usepackage{epstopdf}
\usepackage{tikz}
\usepackage{epstopdf}
\usepackage{mathtools}
\usepackage{enumitem}
\usepackage{subfigure}
\usepackage{tabulary}
\usepackage{color}
\usepackage{url}
\usepackage{flushend}
\usepackage{stfloats}
\usepackage{tcolorbox}
\usepackage[utf8]{inputenc}
\usepackage[english]{babel}
\usepackage{tabularx}
\usepackage{CJKutf8}

\def\BibTeX{{\rm B\kern-.05em{\sc i\kern-.025em b}\kern-.08emT\kern-.1667em\lower.7ex\hbox{E}\kern-.125emX}}

\copyrightyear{2019}
\acmYear{2019}
\acmConference[CIKM '19]{The 28th ACM International Conference on Information and
Knowledge Management}{November 3--7, 2019}{Beijing, China}
\acmBooktitle{The 28th ACM International Conference on Information and Knowledge
Management (CIKM '19), November 3--7, 2019, Beijing, China}
\acmPrice{15.00}
\acmDOI{10.1145/3357384.3358046}
\acmISBN{978-1-4503-6976-3/19/11}

\newcommand{\ModelName}{\textsc{Qrefine}}

\begin{document}
\fancyhead{}
  
\title{Generative Question Refinement with Deep Reinforcement Learning in Retrieval-based QA System}

\author{Ye Liu$^1$, Chenwei Zhang$^{2}$\footnotetext[2]{Work done while at University of Illinois at Chicago}, Xiaohui Yan$^3$, Yi Chang $^4$, Philip S. Yu$^{1}$}
\affiliation{$^1$Department of Computer Science, University of Illinois at Chicago, IL, USA\\
$^2$Amazon, Seattle, WA, USA
$^3$ Poisson Lab, Huawei Technologies, Shenzhen, China \\
$^4$School of Artificial Intelligence, Jilin University, China
}
\email{yliu279@uic.edu, cwzhang@amazon.com, yanxiaohui2@huawei.com, yichang@jlu.edu.cn, psyu@uic.edu}

\begin{abstract}
In real-world question-answering (QA) systems, ill-formed questions, such as wrong words, ill word order and noisy expressions, are common and may prevent the QA systems from understanding and answering accurately. In order to eliminate the effect of ill-formed questions, we approach the question refinement task and propose a unified model, {\ModelName}, to refine the ill-formed questions to well-formed questions. The basic idea is to learn a Seq2Seq model to generate a new question from the original one. To improve the quality and retrieval performance of the generated questions, we make two major improvements: 1) To better encode the semantics of ill-formed questions, we enrich the representation of questions with character embedding and the contextual word embedding such as BERT, besides the traditional context-free word embeddings; 2) To make it capable to generate desired questions, we train the model with deep reinforcement learning techniques that consider an appropriate wording of the generation as an immediate reward and the correlation between generated question and answer as time-delayed long-term rewards. Experimental results on real-world datasets show that the proposed {\ModelName} can generate refined questions with high readability but fewer mistakes than original questions provided by users. Moreover, the refined questions also significantly improve the accuracy of answer retrieval.
\end{abstract}

\newcommand*{\origrightarrow}{}
\let\oldarrow\textrightarrow
\renewcommand*{\textrightarrow}{\fontfamily{cmr}\selectfont\origrightarrow}



\maketitle

\section{Introduction}\label{sec:intro}
QA systems greatly facilitate the information access of users, which are popularly used in both Web and mobile Internet. In these systems, users input a question either by text or voice and expect to get the right answer quickly. However, due to factors such as thoughtless questions from users, misoperations of keyboard input and ASR (automatic speech recognizer) error, the ill-formed questions asked by users are usually expressed with vagueness, ambiguity, noises and errors. 

By manual analysis on the WikiAnswer dataset \footnote{\url{http://knowitall.cs.washington.edu/oqa/data/WikiAnswers}}, we find that about 68\% questions are ill-formed. As shown in Table 1, there are three typical ill-formed question types, specifically, wrong words, ill words order and noisy background, and they include 79\%, i.e., ((21\% + 23\% + 12\%)/68\%) ill-formed questions. Generally, a question is a short sentence with a few words in QA systems. Directly using ill-formed questions to search for answers in a retrieval based QA systems \cite{faruqui2018identifying} will hurt the downstream steps, e.g., answer selection \cite{tan2015lstm} and hence compromise QA systems' effectiveness. 

Inspired by the task of query refinement in web search \cite{nogueira2018learning}, we study the task of question refinement in QA system, which aims to improve the quality of users' questions, in the meanwhile, boost the accuracy of the downstream answer retrieval. We can see that the task is complex since it contains the following subtasks: 1) word correction, e.g., correct \emph{``defenition''} to ``definition''; 2) word reorder, e.g., ill words order example in Table \ref{tab::intro_demo}; 3) sentence simplification, e.g., remove the redundant expression like ``based on tiresias prediction'' in noisy background example in Table \ref{tab::intro_demo}.

\begin{table*}[htb!]
\centering
\resizebox{0.98\textwidth}{!}{%
\begin{tabular}{p{1.5in}|p{0.8in}|p{2.9in}|p{2.1in}}
\hline\hline
\textbf{Type of Ill-formulation} & \textbf{Ratio in Data} & \textbf{Ill-formed Question} & \textbf{Well-formed Questions} \\
\hline
Wrong words & 21\% & \texttt{what is the \emph{\underline{defenition}} of the word infer} &  \texttt{what is the \underline{definition} for inference}\\\hline
Ill words order & 23\% & \texttt{limestone is what kind of \emph{\underline{rocke}}} & \texttt{what is limestone \underline{rock}}\\\hline
Noisy background & 12\% & \texttt{based on tiresias prediction which heroic qualities will odysseus need to rely upon as he continues his journey} & \texttt{what heroic qualities does odysseus rely on} \\ \hline\hline
\end{tabular}
}
\caption{Examples of the three common types of ill-formed and generated well-formed questions on WikiAnswers dataset. The ratio in data is counted within 1000 random sampled triples. The other ill-formed questions belong to several other types which have a minority percentage.}
\label{tab::intro_demo}
\end{table*}

An intuitive way is to tackle these problems one by one alone. For instance, Xie et.al \cite{xie2016neural} proposed a charater-level text correction to deal with the orthographic errors and rare words. Yuan et.al \cite{yuan2016grammatical} focus on grammar error correction to correct the erroneous word phrases. Besides, Zhang et.al. \cite{zhang2017sentence} utilized deep reinforcement learning to simplify questions, like splitting complex questions and substitutes difficult words with common paraphrases. However, it's laboursome to combine these methods together in practice, which might require no domain knowledge and a few human intervention.

Is it possible to tackle these problems with a unified model? Inspired by the successful usage of sequence-to-sequence (Seq2Seq) model \cite{sutskever2014sequence} on related tasks such as machine translation \cite{britz2017massive}, text summarization \cite{nallapati2016abstractive}, and sentence simplification \cite{zhang2017sentence}, it is promising to use it in the question refinement task. Seq2Seq model is flexible enough to encode patterns for sequence transformation such as word correction, word recorder, and sentence simplification, if there are appropriate training datasets. Unfortunately, we find that the vanilla Seq2Seq model does not perform well on this task. The reasons may be twofold: 1) it fails to learn a good representation of ill-formed questions, which might contain many wrong or noisy words. 2) The maximize likelihood objective is not consistent with our target, i.e., generated better quality questions and thus improve the accuracy of answer retrieval.

To overcome these problems, we develop a Seq2Seq model for question refinement called QREFINE. For the question representation, since a well-formed question might sensitive to the word order, we make use of the recent proposed contextual word embeddings such as BERT \cite{devlin2018bert} to capture the contextual word information. As BERT is trained over a large scale unlabeled corpus, it also can alleviate the data sparsity problem where there is not enough training data. Moreover, considering the ill-formed questions might contain typos, we also incorporating the fine-grained character embedding \cite{pan2017memen} as a part of question representation. Our experimental results show that the two types of representations substantially improve the effectiveness of the Seq2Seq model. 

To make the Seq2Seq model generate desired questions, we develop a training algorithm based on reinforcement learning. we assign not only word-level rewards to each word for its wording from a pertained language model and Bert language model as immediate rewards but also question-level rewards such as the correlation of the refined question to its answer. In order to solve the low data efficiency and unrobust policy problems on the traditional policy gradient method, we use advanced policy gradient method proximal policy optimization (PPO) \cite{tuan2018proximal} for well-formed question generation \cite{schulman2017proximal, tuan2018proximal}. We compared our model with the state-of-the-art baselines in two QA datasets. The result shows our model outperforms baselines on question refinement. Besides, the case studies show the improved readability of the questions after refinement using {\ModelName}, and its effectiveness in improving the utility of an existing QA system. Moreover, it's worth to notice that our model is fully data-driven and might not require domain knowledge and human intervention.

\section{Preliminary}\label{sec:Pre}
We formally define the question refinement task studied in this paper. After that, we introduce some terminologies that we will use throughout the paper.

\subsection{Problem Description}
Given an ill-formed question consists of $\mathbf{x} = [x_{1}, x_{2}, ..., x_{N}]$ of an arbitrary-length $N$, the well-formed question $\mathbf{y} = [y_{1}, y_{2}, ..., y_{M}]$ of a variable-length $M$. The aim of question refinement is to refine $\mathbf{x}$ to $\mathbf{y}$ which has better readability. It is expected that the generated well-formed question $\mathbf{y}$ can be better able to retrieve the best answer candidate $\mathbf{a_k} = [a_{1}, a_{2}, ..., a_{L}]$, where $1 \leq k \leq s$ from an answer candidate pool $\{\mathbf{a}_1, \mathbf{a}_2, ..., , \mathbf{a}_s\}$. 

\subsection{Seq2Seq Framework on Question Refinement}
The Seq2Seq model adopts an encoder-decoder framework that learns to encode an ill-formed question $\mathbf{x}$ into a fixed-length vector representation and to decode the fixed-length vector back into a variable-length well-formed question $\mathbf{y}$. From a probabilistic perspective, Seq2Seq model is a general method that learns the conditional distribution over a variable-length sequence conditioned on another variable-length sequence, namely, $p_{lm}(y_{1}, ..., y_{M} | x_{1}, ..., x_{N})$. 

The encoder can be a convolution neural network or a recurrent neural network that summarizes the ill-formed question into a vector representation. Since LSTM \cite{hochreiter1997long} is good at learning long-term dependencies in the data \cite{graves2013generating}, we adopt LSTM to sequentially encode each word of ill-formed question $\mathbf{x}$. As the LSTM  reads each word, the hidden state of the LSTM is updated $h_n={\rm LSTM_{\rm encoder}}(h_{n-1},x_{n})$. Therefore, the encoder transforms the ill-formed question $\mathbf{x}$ into a sequence of hidden states $(h_{1}, h_{2}, ..., h_{N})$.

The decoder can be another LSTM which is trained to generate the current hidden state $k_{m}$ based on the current word $y_{m}$ and the previous hidden state $k_{m-1}$:
\begin{equation}
k_{m} = {\rm LSTM}_{\rm decoder}(k_{m-1}, y_{m}).    
\end{equation}

Moreover, as introduced in \cite{luong2015effective}, a context vector $c_m$ can be obtained for each decoder step $m$ by being attentive to the encoding of the source question dynamically:
\begin{align}
    c_{m}=\sum_{n=1}^{N} \alpha_{nm} h_{n},
\end{align}
where $c_m$ is a weighted sum of the hidden states of the ill-formed question $\mathbf{x}$:
The attention score $\alpha_{nm}$ between the $n$-th ill-formed question hidden unit $h_{n}$ and $m$-th well-formed question hidden unit $k_{m}$ is calculated as follows: 
\begin{align}
\alpha_{nm} = \frac{exp(h_{n}^{T} \cdot k_{m})}{\sum_{l=1}^{N} exp(h_{l}^{T} \cdot k_{m})}.
\end{align}
Formally, the Seq2Seq model is formed as below:
\begin{align} \label{softmax}
    p_{lm}(y_{m}| y_{1:m-1}, \mathbf{x}) = {\rm softmax}(g(k_{m}^{T}, c_m)),
\end{align}
where $g(\cdot)$ is an activation function $g(k_{m}^{T}, c_m) = \textbf{W}_{o} {\rm tanh}(\textbf{U}_{h}k_{m}^{T} + \textbf{W}_{h}c_{m})$, where $\textbf{W}_{o}\in \mathbb{R}^{|V|\times d}, \textbf{U}_{h} \in \mathbb{R}^{d \times d}$ and $\textbf{W}_{h} \in \mathbb{R}^{d \times d}$; $|V|$ is the output vocabulary size and $d$ is the hidden unit size.

\begin{figure*}[t]
\centering
\includegraphics[width=1.0\linewidth]{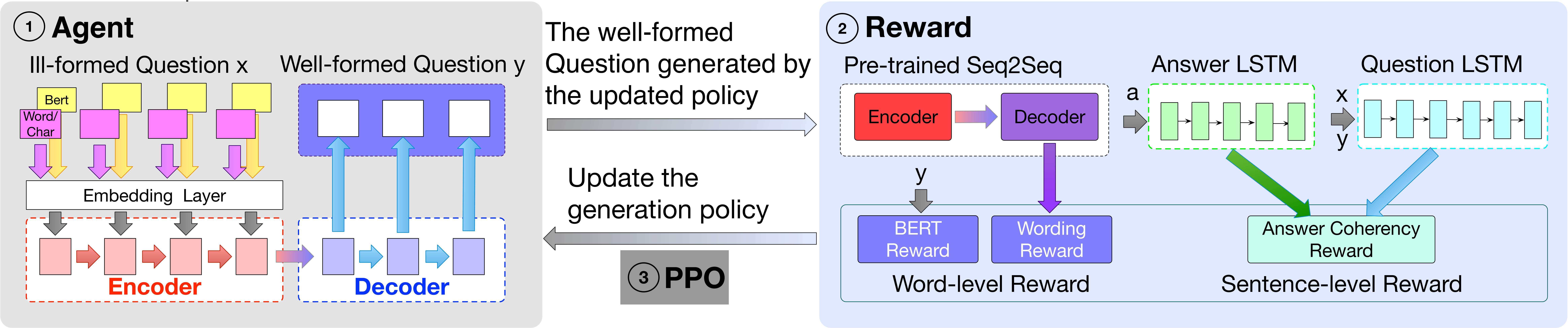}
\caption{The architecture of the proposed model {\ModelName}. \textcircled{1} The encoder of the agent module reads the ill-formed question and the decoder generates a well-formed question, one word/phrase at a time. \textcircled{2} The well-formed question being generated so far is sent to a pre-trained reward module, which calculates a word-level wording reward from word-level LM and BERT Reward and a question-level answer correlation reward from QA similarity. \textcircled{3} The PPO module updates agent's generation policy, aiming to maximize the rewards.}
\label{fig:Frame}
\end{figure*}

\section{Reinforced Generative Question Refinement}\label{sec:RLQR}
\subsection{Model Description}
Despite the successful application in numerous sequence transduction tasks \cite{bahdanau2014neural}, a vanilla Seq2Seq model is not ideal for question refinement since it only makes a few trivial changes of ill-formed question \cite{zhang2017sentence}. To encourage a wider variety of rewrite operations while keeping the refined question fluent and coherent to the answer, we employ a reinforcement learning framework (see Figure \ref{fig:Frame}). 

The refinement agent first reads the ill-formed question $\mathbf{x}$ from the encoder; and then at each step of the decoder, it takes an action $y_{m} \in V$, where $V$ is the output vocabulary, according to a policy $\pi_{\theta}(y_{m}|y_{1:m-1}, \mathbf{x})$. The agent continues to take actions until it produces $<$EOS$>$ (denoting end of sentence) token yielding the generated well-formed question of our model $\mathbf{y} = [y_{1}, y_{2}, ..., y_{M}]$. Two types of rewards, wording reward, and answer correlation reward, are received and the advanced policy gradient method PPO is used to update the agent. In the following, we introduce our question representation and reward function. After that, we present the details of the processes for generating accurate and consistent questions by leveraging the REINFORCE and PPO method.

\subsection{Question Representation}
The ill-formed question has a wrong semantic order and contains some unrelated background information. To solve the problem, the model need to learn the institutional utterance of words, which needs to consider the correlation between words and words. The widely-used context-free models such as Skip-gram \cite{mikolov2013distributed} or GloVe \cite{pennington2014glove} cannot consider the correlation between words. Because they generate a single word embedding representation for each word in the vocabulary, so ``apple'' would have the same representation in ``red apple'' and ``apple store''. However contextual models, like BERT can generate a representation of each word based on the other words in the sentence. Therefore, we concatenate the context-free and contextual embedding together as the word embedding to capture such coarse-grained correlation patterns of words.

Since the ill-formed question always contains the misspelled words, which is usually set as $<$UNK$>$, which is hard to capture the meaning of the original word. To capture the meaning of the misspelled words, we extend the word expression by incorporating fine-grained character expression. By using the character-level embedding, we can get the high-dimensional vector representation. As character-level input, the original sentence is decomposed into a sequence of characters, including special characters, such as quotation mark. Characters are embedded into vectors, which can be considered as 1D inputs to the Bidirectional Long Short-term Memory Network (BI-LSTM) \cite{hochreiter1997long}, and the hidden layer of the last LSTM unit is the fixed-size vector of each word. Overall, we combine the fine-grained character-level embedding and coarse-grained contextual and context-free word embedding to represent the question. 
 
\subsection{Reward}\label{sec:reward}
The reward for system output $\mathbf{y}$ is the weighted sum of two types of rewards aimed at achieving well-formed readable and answer correlation question: wording rewards on the word-level from the Reward RNN and BERT, which aims to measure how well each generated word is in line with the language model (LM) rule, and the question-level answer correlation reward that has the ability to infer the correlation of the refined question to its answer, even if it is not generating until the end of the well-formed question. 

~\\
\textbf{Wording Reward} 
The wording reward $r_{w}$ aims to give an immediate reward to each of words when it is being generated in the well-formed question. 
BERT pre-trained on the large dataset like Wikipedia could give the contextual wording reward $r_{B}(y_t) = p_{B}(y_{t}|y_{1,..., M})$, which is the probability of word $y_{t}$ given by BERT model. Moreover, for the domain-specific, we also use the decoder of the pre-trained Seq2Seq module as a trained LM Reward RNN which is able to score the probability of the next word given the words generated so far. Thus, the wording reward of the $t$-th word in the well-formed question is:
\begin{align}
    r_{w}(y_{t})= r_{B}(y_{t}) + p_{lm}(y_{t+1}|k_{t}),
\end{align}
where $k_{t}$ is the current state which is the hidden representation of the generated well-formed question with $t$ words so far: $[y_{1}, y_{2}, ..., y_{t}]$ and $y_{t+1}$ is the generated word in the $(t+1)$-step.

~\\
\textbf{Answer Correlation Reward}
The refinement result should not only improve the readability of the question, more importantly, have better ability to address its correlation to the answer once refined. With this motivation, we design an answer correlation reward $r_{ac}$ to further measure the correlation of the refined question to its answer on the question-level as a whole. As answers themselves are sometimes ill-formed and contain a large amount of unrelated information, they may not share lexical units with the well-formed question directly. But the well-formed question is semantically easier to answer than the ill-formed question. Following the similar ranking loss in \cite{tan2015lstm, feng2015applying}, the answer correlation module defines the training objective as a hinge loss:

\begin{align}
  & r_{ac}(\mathbf{y})  =  max\{ 0,  \epsilon - \rm sim(\rm LSTM_q(\textbf{x}),   \\ \nonumber
  &  \rm LSTM_a(\textbf{a})) + \rm sim(\rm LSTM_q(\textbf{y}), \rm LSTM_a(\textbf{a})) \},
  \label{ans_rewd}
\end{align}
where we use two separate \small $\rm LSTM$\normalsize s,  \small $\rm LSTM_q$ \normalsize and \small $\rm LSTM_a$\normalsize, to encode the question and answer to the vector representation. The well-formed question and ill-formed question share the same \small $\rm LSTM_q$ \normalsize, and $\epsilon$ is the constant margin. Furthermore, \small $\rm sim(\rm LSTM_q(\textbf{x}),\rm LSTM_a(\textbf{a})) = \rm LSTM_q(\textbf{x})\mathbf{W}_{\rm sim} \rm LSTM_a(\textbf{a})^{T}$ \normalsize computes a bi-linear term between the question and its correct answer. We train the model by maximizing answer correlation reward $r_{ac}$ using ground-truth well-formed and ill-formed questions to learn the weight of \small $\rm LSTM_q$ \normalsize, \small $\rm LSTM_a$ \normalsize network and ${\mathbf{W}}_{\rm sim}$. After that, we hold a fixed copy of networks to give rewards to the generated well-formed question.  

~\\
\textbf{Accumulated Reward}
We add the answer correlation reward to the end of the wording reward, as the overall evaluation of the generated question. The {\ModelName} reward $r$ of each word is the combination of the wording reward and the answer correlation reward,
\begin{align}
    r(y_i) = \left\{  
    \begin{array}{lr}  
             r_{w}(y_i), i \neq M \\  
             r_{w}(y_i) + c_{1} r_{ac}(\mathbf{\textbf{y}}), i = M \\  
    \end{array}  
    \right.
\end{align}
where $c_{1}$ is the parameter to tune the weight between wording reward $r_{lm}$ and answer correlation reward $r_{ac}$;
Since we want {\ModelName} module to have the ability to infer reward even if not reaching the end of the generation process and the future reward will influence the current reward, we adopt the accumulated {\ModelName} reward $R$ with the discounted factor $\gamma$ and the accumulated {\ModelName} discounted reward of $t$-th word is represented as, 
\begin{equation}
    {R}(y_{t}) =  \gamma^0 r(y_{t}) + \gamma^1 r(y_{t+1}) + \dots + \gamma^{M-t} r(y_{M}).
    \label{accumulate}
\end{equation}
By using the accumulated reward, we are able to infer the answer correlation reward even if we are not reaching the end of the generation process.

\subsection{Question Generation} \label{sec:model}
A popular choice of loss in traditional models is the cross-entropy used to maximize the probability of the next correct word. However, this loss is at the word-level and the performance of these models is typically evaluated using discrete metrics. To overcome, we draw on the insights of deep reinforcement learning, which integrates exploration and exploitation into a whole framework. Instead of learning a sequential recurrent model to greedily look for the next correct word, we utilize a policy network and a reward function to jointly determine the next best word at each time step, which aims to maximize the reward of the whole sentence.

Question refinement can be formulated as a Markov decision process (MDP) $(S, A, P, R)$, where $S$ is a set of states $s_t = \{y_{1: t-1}, \textbf{x}\}$, $A$ is a set of actions $a_{t} = y_{t}$, $P$ is the transition probability of the next state given the current state and action, $R$ is a reward function $r(s_{t}, a_{t})$ for every intermediate time step $t$, and $\gamma$ is a discount factor that $\gamma \in [0, 1]$.
 
The actions are taken from a probability distribution called policy $\pi$ given the current state (i.e., $a_t \sim \pi(s_t)$). In question refinement, $\pi$ is a seq2seq model. Therefore, reinforcement learning methods are suitable to apply to question refinement model by learning the seq2seq model, or policy $\pi$, that can gain reward as much as possible.

\subsubsection{On-policy Optimization}
Due to the high dimensional action space for question refinement and high diversity of the required generation result, policy gradient method, like REINFORCE \cite{barto1998reinforcement} are more appropriate in the question generation than value-based methods like Q-learning \cite{mnih2013playing}.

For a given ill-formed question $\mathbf{x}$, we want to return a formulated question $\mathbf{y}$, maximizing an accumulated reward $R$. The answer $\mathbf{a}$ is the known given by the database. The question $\mathbf{y} \sim \pi_{\theta}(\cdot|\mathbf{x})$ is generated according to $\pi_{\theta}$ where $\theta$ is the policy's parameter and the goal is to maximize the expected reward of the reformulated question under the policy, $E_{\mathbf{y} \sim \pi_{\theta}(\cdot|\mathbf{x})}[R(\mathbf{y})]$.

Given reward $r_t$ at each time step $t$, the parameter $\pi$ of policy $\pi$ (a seq2seq model) is updated by policy gradient as follows:
\begin{align}
    E_{\mathbf{y} \sim\pi_{\theta}(\cdot|\mathbf{x})} [R(\mathbf{y})]\approx \frac{1}{N} \sum_{i=1}^{N} R(y_i), y_i \sim \pi_{\theta}(\cdot|\mathbf{x}).
\end{align}
To compute gradients for training we use REINFORCE\cite{williams1991function}, 
\begin{align}
    & \nabla  E_{\mathbf{y} \sim \pi_{\theta}(\cdot|\mathbf{x})} [R(\mathbf{y})]\\ \nonumber
    &  = E_{\mathbf{y} \sim \pi_{\theta}(\cdot|\mathbf{x})} \nabla_{\theta} log(\pi_{\theta}(\mathbf{y}|\mathbf{x})) R(\mathbf{y}) \\ \nonumber
    & \approx \frac{1}{N} \sum_{i=1}^{N} \nabla_{\theta} log(\pi_{\theta} (y_i|y_{1:i-1}, X)) R(y_i), y_i \sim \pi_{\theta}(\cdot|\mathbf{x}).
\end{align}

This estimator is often found to have high variance, leading to unstable training \cite{greensmith2004variance}. We reduce the variance by using a baseline: $B(\mathbf{x}) = E_{\mathbf{y} \sim \pi_{\theta}(\cdot|\mathbf{x})} [R(\mathbf{y})]$ \cite{sutton2000policy}. This expectation is also computed by sampling from the policy given $\mathbf{x}$.

To avoid the model not being able to explore new words that could lead to a better answer correlation question, we use entropy regularization:

\begin{align}
    \mathbf{H}[\pi_{\theta}(\mathbf{y}|\mathbf{x})] = \sum_{i=1}^{N} \sum_{y_{i} \in V} log(\pi_{\theta}(y_{i}| y_{1:i-1}, \mathbf{x}))\pi_{\theta}(y_{i}| y_{1:i-1}, \mathbf{x})
\end{align}

The final objective is:
\begin{align}
    \mathbb{E}_{\mathbf{y} \sim \pi_{\theta}(\cdot|\mathbf{x})}[R(\mathbf{y}) - B(\mathbf{x})] + \lambda \mathbf{H}[\pi_{\theta}(\mathbf{y}|\mathbf{x})],
\end{align}

where $\lambda$ is the regularization weight. $R(\mathbf{y}) - B(\mathbf{x})$ can be interpreted as the goodness of adopted action at over all the possible actions at state st. Policy gradient directly updates $\pi$ to increase the probability of at given $s_t$ when advantage function is positive, and vice versa.

\subsubsection{Off-policy Optimization}
The vanilla policy gradient method is on-policy method and have convergence problem \cite{schulman2015trust}. Empirically they often lead to problems like low data efficiency and unreliable performance, as shown in subsection \ref{sub:lc}. We use the advanced deep reinforce method proximal policy optimization (PPO) \cite{schulman2017proximal} to learn a more stable policy.

Proximal policy optimization (PPO) \cite{tuan2018proximal} is an approximation method of trust region policy optimization (TRPO) \cite{schulman2015trust}. Different from TRPO which uses a second-order Taylor expansion, PPO uses only a first-order approximation, which makes PPO very effective in RNN networks and in a wide distribution space:

\begin{align}
   & max_{\theta}L^{TRPO}(\theta) = \mathbb{E}[\frac{\pi_{\theta}(a_{t}|s_{t})}{\pi_{\theta_{old}(a_{t}|s_{t})}} A_{t}], \\ \nonumber
   & \text{subject to } \mathbb{E}[\text{\textbf{KL}}[\pi_{\theta_{old}(a_{t}|s_{t})}, \pi_{\theta(a_{t}|s_{t})}]] \leq \delta
\end{align}

where $\pi_{old}$ is the old parameters before update. The updated policy $\pi$ cannot be too far away from the old policy $\pi_{old}$, because the KL-divergence between $\pi$ and $\pi_{old}$ is bounded by the small number $\delta$.

To optimize policy, PPO alternates between sampling sentence generated from the current policy and performing several epochs of optimization on the sampled sentences. According to the paper \cite{schulman2017proximal}, the clipped objective to heuristically constrain the KL-divergence setting achieves the best performance:
\begin{align}
    L_{t}^{CLIP}(\theta) = \mathbb{E}_{t}[min(r_{t}(\theta)
    clip(r_{t}(\theta), 1-\epsilon, 1 + \epsilon)) \hat{A}_{t}],
\end{align}

where $\beta_{t}$ denotes the probability ratio $\frac{\pi_{\theta}(a_{t}|s_{t})}{\pi_{\theta_{old}}(a_{t}|s_{t})}$ and $\epsilon$ is a hyperparameter (e.g., $\epsilon = 0.1$). When $\hat{A}_{t}$ is positive, the objective is clipped by $(1 + \epsilon)$; otherwise the objective is clipped by $(1 - \epsilon)$.

$\hat{A}_{t}$ is the expected advantage function (the expected rewards minus a baseline like value function $V(k_t)$ of time $k_t$) which can be calculated as:
\begin{align}\label{advantage}
    &\hat{A}_{t} = \delta_{t} + (\gamma\lambda)\delta_{t+1} + \cdots + \cdots + (\gamma\lambda)^{T-t+1} \delta_{T-1},\\ \nonumber
    & \delta_{t} = r_{t} + \gamma V(k_{t+1}) - V(k_{t}).
\end{align}

To improve the exploration of our model for generating diverse yet coherent words that could constitute a better well-formed question, we use entropy regularization. The integrated PPO method is shown as below: 
\begin{equation}
    L_{t}^{PPO}(\theta) = \mathbb{E}_{t}[L_{t}^{CLIP}(\theta)+ c_{2}\pi_{\theta}(y_{t+1}|k_{t})log(\pi_{\theta}(y_{t+1}|k_{t}))].
    \label{PPo}
\end{equation}
The algorithm of QREFINE with PPO optimization shows in Alg. \ref{algo:qrefine}

~\\
\textbf{Learning} 
The training stage of traditional models suffer from the exposure bias \cite{ranzato2015sequence} since in testing time the ground-truth is missing and previously generated words from the trained model distribution are used to predict the next word. In the testing phase, this exposure bias makes error accumulated and makes these models suboptimal, not able to generate those words which are appropriate but with low probability to be drawn in the training phase.

In order to solve the exposure bias problem, we train the model by using MIXER algorithm described in \cite{ranzato2015sequence} to expose both training data and its predictions. In the inference stage, we greedily selected the word that has the highest probability to generate the question stopping until the $<$EOS$>$ token is generated. 

\renewcommand{\algorithmicrequire}{\textbf{Input:}}
\renewcommand{\algorithmicensure}{\textbf{Output:}}
\begin{algorithm}[t]
\caption{QREFINE-PPO}
\begin{algorithmic}[1]
\REQUIRE Ill-formed question $X$, Well-formed question $Y$, rating data $R$, the number of episodes $K$,  $\varepsilon$-greedy parameter $\varepsilon$,
ratio of language model reward and answer correlation reward $c_1$, the discounted factor of RL $\lambda$, the threshold $\epsilon$ and the entropy regularization $c_2$
\ENSURE  the learned policy $p_{\theta}$
\STATE Initialize policy $p_{\theta}$ and old policy $p_{\theta_{old}}$ with supervised pretrained policy $p_{\theta'}$
\FOR{episode = 1, ..., K} 
\STATE {Uniformly pick a batch of ill-formed question $u \in \mathcal{U}_{train}$ as the environment}
\STATE {Start to generate the word according to $p_{\theta_{old}}(y_{i}|X)$ until the $<$EOS$>$ token is generated, the generated sentence as $Y'$}
\STATE {Send $X$ and $Y'$ to the BERT mechine and pretrained word embedding model to calculate the word-level reward}
\STATE {Send $X$, $Y$ and $Y'$ to the qa-lstm model to calculate the sentence-level reward, based on Eq. \ref{ans_rewd}}
\STATE {Calculate the advantage function of each time step according to Eq. \ref{advantage}}
\REPEAT
\STATE {Update the policy $p_{\theta}$ using Eq. \ref{PPo}}
\UNTIL{convergence}
\STATE {Set old policy $p_{\theta_{old}}$ to policy $p_{\theta}$}
\ENDFOR
\end{algorithmic}
\label{algo:qrefine}
\end{algorithm}

\section{Experiments}\label{sec:exp}
In this section, we evaluate the proposed methods on two real-world datasets, by examining the readability of the refined questions both quantitatively and qualitatively. A question-answer retrieval experiment is used to evaluate the performance of the generated question. In the end, we further conduct the ablation study and learning curve of the method.

\subsection{Dataset}
The datasets are formatted as triples, where each instance consists of a well-formed question, an ill-formed question and an answer in each triple. \\
\textbf{Yahoo}: The Yahoo non-factoid question dataset \footnote{\url{https://ciir.cs.umass.edu/downloads/nfL6/}} is collected from Yahoo Webscope that the questions would contain non-factoid answers in English. After limiting the length of the answer, we have 85k questions and their corresponding answers. Each question contains its best answer along with additional other answers submitted by users. The average length of ill-formed questions is around 12 tokens and the average length of well-formed questions is around 10 tokens, with 73k vocabulary size and 44 different characters. The average length of answers is around 39 tokens. 

To testify the refinement performance from ill-formed to well-formed question, we generate the ill-formed question on three types of the ill-formed question, wrong words, wrong order and Noisy background. We randomly change the character of the words or change the order of the character of words to generate the \textbf{Wrong Word} dataset. 
For the \textbf{Wrong Order} dataset, we randomly change the order of some fragments in the well-formed question. For the \textbf{Noisy Background} dataset, we randomly sampled an arbitrary length phrase from the other answer and add to the original clean question. For the \textbf{Yahoo} dataset, we randomly execute those three operations to generate threefold noisy questions, which contains 254k triples. The example of ill-formed question in those datasets are shown in Table \ref{tab:three_operations}.

\begin{table*}[htb!]
\resizebox{0.85\textwidth}{!}{
\begin{tabular}{l|l}
\toprule[1.25pt]
\multicolumn{2}{c}{\textbf{Original well-formed question : why don't european restaurants serve water?}} \\
\hline 
\textbf{Type}              & \textbf{Ill-formed question} \\
\hline
Wrong word/typo   & \texttt{why don't \emph{\underline{oeurpan rantaurest}} serve \emph{\underline{wataar}}?}                 \\
\hline
Wrong order      &  \texttt{european restaurants why serve don't water?}       \\
\hline
Noisy background &  \texttt{concerned with the digestive process why don't european restaurants serve water?}           \\
\hline
Three operations & \texttt{why \emph{\underline{digistive}} process with the \emph{\underline{restarunts}} european don't serve water?} \\
\bottomrule[1.25pt]
\end{tabular}
}
\caption{Example of the three operations to generate ill-formed question on Yahoo Dataset}
\label{tab:three_operations}
\end{table*}

\textbf{Customer Service Userlog (CSU)}: This anonymized dataset contains online question-answering userlog from a commercial customer service platform containing 1 million instances in chinese language. The ill-formed question is the question asked by users and the well-formed question is selected from a pool of FAQs collected by editors. After we delete the duplicated triples, the left triple size is 111k. The average length of ill-formed questions is around 6 tokens, and the average length of well-formed questions is also around 6 tokens while the average length of answers is around 54 tokens with 14k vocabulary size and 2041 characters. \\
\subsection{Baselines and Benchmarks}
The compared methods are summarized as follows:
\begin{itemize}
    \item \textbf{Seq2Seq} is a basic encoder-decoder sequence learning system with Luong attention \cite{luong2015effective} and Bi-direction LSTM on encoder model. 
    \item \textbf{PARA-NMT} \cite{dong2017learning} is a NMT-based question paraphrasing method which assigns higher weights to those linguistic expressions likely to yield correct answers. 
    \item \textbf{AQA} \cite{buck2018ask} is the reinforce method seeking to reformulate questions such that the QA system has the best chance of returning the correct answer in the reading comprehension task. Since our datasets do not contain the context information, we use the QA-lstm to measure the similarity between the generated question and the answer as the reward. Following \cite{britz2017massive}, we use a bidirectional LSTM as the encoder and a 4-layer stacked LSTM with attention as the decoder. 
    \item \textbf{TOQR} \cite{nogueira2017task} is the query reformulation method with reinforcement learning to maximize the number of relevant documents returned. TOQR use reinforcement method to select terms from the original query and candidate retrieved document to reformulate the query, and the reward is the document recall.
\end{itemize}

Since the proposed {\ModelName} consists of several components, we consider several variations of {\ModelName} as follows: \\
\textbf{QR-word} is the reinforce model only using wording reward, which is viewed as word-level reward. 
\textbf{QR-ans} is the reinforce model using answer correlation as the reward, which is views as question-level reward.
\textbf{{\ModelName-RF}} combines both word-level wording reward and question-level answer correlation reward and uses REINFORCE policy gradient based to optimize.\\
\textbf{{\ModelName-PPO}} is the proposed model using PPO and combining both word-level wording reward and question-level reward. \\
The code is available on github \footnote{\url{https://github.com/yeliu918/QREFINE-PPO}} \\

\begin{table*}
\centering
\resizebox{0.85\linewidth}{!}{
\begin{tabular}{c|ccc|ccc|ccc}
\toprule[1.25pt]
                         &                                 & \textbf{Wrong Word}                              &                                 &                                 & \textbf{Wrong Order}                              &                                 &                                 & \textbf{Noisy Background}                              &                                 \\ \cline{2-10}
\textbf{Method}                                                  & \textbf{BLEU-1}                          & \textbf{Rouge}                           & \textbf{Meteor}                         & \textbf{BLEU-1}                          & \textbf{Rouge}                          & \textbf{Meteor}                         & \textbf{BLEU-1}                          & \textbf{Meteor}                          & \textbf{Rouge}                          \\ \hline
Seq2Seq & 47.10                           & 60.41                           & 30.95                           & 53.11                           & 67.67                           & 36.37                           & 50.17                           & 62.19                           & 31.75                           \\ 
PARA-NMT & 53.10                           & 64.91                           & 35.59                           & 59.13                           & 73.75                           & 41.27                           & 55.37                           & 68.45                           & 37.56                           \\ 
TOQR                        & 43.15                           & 56.04                           & 28.47                           & 49.60                           & 45.49                           & 56.76                           & 32.75                           & 57.85                           & 45.39                           \\ 
AQA                           & 61.93                           & 77.36                           & 45.91                           & 63.34                           & 80.83                           & 50.15                           & 61.08                           & 74.14                           & 42.10                           \\ \hline
QREFINE-RF                  & 67.82                           & 83.16                           & 51.07                           & 70.74                           & 87.21                           & 55.24                           & 69.12                           & 85.21                           & 53.17 \\
QREFINE-PPO & \textbf{68.83} & \textbf{84.76} & \textbf{52.72} & \textbf{72.22} & \textbf{88.94} & \textbf{56.22} & \textbf{71.57} & \textbf{86.12} & 
\textbf{53.22} \\  
\bottomrule[1.25pt]
\end{tabular}
}
\caption{Question Generation Evaluation on Yahoo dataset to test models ability to correct wrong words, order and remove background.}
\label{Yahoo-three}
\end{table*}

\begin{table*}
\centering
\resizebox{0.85\linewidth}{!}{
\begin{tabular}{l|cccccc|cccccc}
\toprule[1.25pt]
& & & \textbf{Yahoo}  & & & & &  & \textbf{CSU}  & &\\ \cline{2-13}
 \textbf{Method}   & \textbf{BLEU-1} & \textbf{BLEU-2} & \textbf{BLEU-3} & \textbf{BLEU-4} & \textbf{Rouge} & \textbf{Meteor} & \textbf{BLEU-1} & \textbf{BLEU-2} & \textbf{BLEU-3} & \textbf{BLEU-4} & \textbf{Rouge} & \textbf{Meteor} \\
\hline
Seq2Seq           & 39.50 &   31.53 &  23.81 &  16.78 &  53.07 &  22.76  & 72.76 &  53.17 &  37.49 &  26.17 & 68.39 &  35.57\\
PARA-NMT &41.08& 33.74 &26.50 &19.73 &55.10  & 23.80 & 73.18 & 59.41 & 56.53 & 36.44 & 69.05 & 57.18\\
{AQA}     &   43.40    & 37.14  & 30.80  &24.58  & 58.37  & 27.17  & 74.13  & 65.53  & 58.00  & 31.63 & 74.40 & 62.67   \\
TOQR         & 31.23 &  20.69 &  12.45 &  5.89 &   41.92 &   15.19 &48.92 &  44.69 &40.34 &35.73 &65.10 &33.00 \\
\hline
QR-word      & 47.73 &  42.46 &  37.09 &  31.46 &  63.03 &  31.00 &  77.49 &   69.61 &  62.00 &   34.91 &  77.71 & 66.17 \\     
QR-ans          & 47.17 &   41.80 &  36.36 &  30.66 &  62.48 &  30.55  & 78.50 &   70.95 &   63.08 &  36.99 &   78.72 &   67.07 \\
{\ModelName}-RF & 48.72 &   44.20 &  39.58 &   34.74 &  64.60 &  32.59 & 79.71 &   72.43 &  64.59 &   37.54 &  79.96 &   68.40\\
{\ModelName}-PPO  & \textbf{50.90}  & \textbf{47.47}  & \textbf{43.91}  & \textbf{40.19} & \textbf{67.41} & \textbf{35.37} & \textbf{82.55}  & \textbf{75.54}  & \textbf{67.63} & \textbf{40.33} & \textbf{82.74} & \textbf{71.54}  \\
\bottomrule[1.25pt]
\end{tabular}
}
\caption{Question Generation Evaluation on Yahoo and CSU dataset.}\label{bleu_score}
\end{table*}

\noindent\textbf{Experimental Setting}
We randomly divide the dataset into a training set ($80\%$), a development set ($10\%$), and a test set ($10\%$). We tune the hyperparameters on development set and report results on test set. We implement all model by Tensorflow using Python on a Linux server with Intel Xeon E5-2620 v4 CPU and an NVIDIA 1080Ti GPU. If the paper were accepted, we would publish code and data online.

On Yahoo dataset, we use the released skip-gram model word embedding \cite{mikolov2013distributed} with 300 dimensions \footnote{\url{https://github.com/facebookresearch/fastText/blob/master/pretrained-vectors.md}}. We fix the word representations during training. The character embedding of is 50 dimensions for each character. The number of hidden unit in character Bi-LSTM is 100, so the size of word through character embedding is 200. We add $<$EOS$>$ at the end of sentences. And we set the word out of the vocabulary to $<$UNK$>$. 

We choose word embedding of 200 dimensions for CSU dataset and use the Glove model \cite{pennington2014glove} to get the pre-trained word embedding. The character embedding of CSU dataset is 50 dimensions for each character. The number of hidden unit in character Bi-LSTM is 50, so the size of word through character embedding is 100. 

For BERT word embedding \footnote{\url{https://github.com/hanxiao/bert-as-service}}, it gives us 768 dimensions of the word embedding on both datasets. We combine contextual-free word embedding, BERT embedding and character embedding for each of word on both dataset. 

We set the LSTM hidden unit size to 300 on CSU dataset and 500 on WikiAnswer dataset. Optimization is performed using Adam \cite{kingma2014adam}, with an initial learning rate of $0.001$; the first momentum coefficient was set to $0.9$ and the second momentum coefficient to $0.999$. The mini-batch size for the update is set at 64 on both datasets. During decoding, we greedily pick the generated word. Decoder stops when the $<$EOS$>$ token is being generated. During reinforcement training, we set the ratio of language model reward and answer correlation reward $c_1$ as $\{0.1, 1, 10\}$, the discounted factor of RL $\lambda$ is $[0, 1)$, the threshold $\epsilon$ is $\{0.1, 0.2, 0.3\}$ and the entropy regularization $c_2$ is $\{0.1, 1\}$. All hyperparameters of our model are tuned using the development set.\\

\subsection{Question Generation}
In this section, we give the experimental analysis to quantitatively and qualitatively evaluate the quality of generated questions. 

\subsubsection{Quantitative Evaluation of Question Generation}
To evaluate the quality of the generated well-fined question, we first use automatic metrics to quantitatively show the performance. We use the precision-based automatic metrics BLEU-1, BLEU-2, BLEU-3, BLEU-4 \cite{papineni2002bleu} which measures the average n-gram precision on a set of reference sentences, with a penalty for overly short sentences, and ROUGE \cite{lin2004rouge} based on recall and METEOR \cite{banerjee2005meteor} that is based on both precision and recall to measure the generation results. 

In Table \ref{Yahoo-three}, our model performs best on each single task, wrong word, wrong order or removing noisy background. Since {\ModelName} uses character-level embedding, contextual-free and contextual word embedding BERT, it can better deal with the misspellings and understand the ill-formed question than all baselines. By using the word-level reward, our model can learn a better language policy, hence it can perform well on correcting the word order and wrong word. Besides, since our model considers the answer correlation as the reward, so it can capture the useful information in the question and achieves superior results on the noisy background. 

For the three operations composite task, as shown in Table \ref{bleu_score}, our model performs the best on both datasets. In PARA-NMT, the paraphrased questions are very similar, and therefore there is a great chance that paraphrased questions always get the same answer. The reinforced query reformulation method TOQR reformulates the query by selecting terms from the retrieved documents by using the document recall as reward. Since the query is a list of terms, therefore, the generated sequence of terms have poor readability, resulting in lowest performance among all baselines. The original AQA refines the question for the reading comprehension. We use the QA-lstm as reward, which has similar framework as PARA-NMT but using REINFORCE optimization. Therefore AQA has the same problem as PARA-NMT. Compared to QR-ans, QR-word has a better performance, which shows that QR-word contributes more to the readability. {\ModelName}-PPO is higher than {\ModelName}-RF, which may because PPO method can improve the high variance problem using REINFORCE method on question refinement task.

\begin{table*}
\centering
\resizebox{0.95\textwidth}{!}{
\begin{tabular}{l|l|l} 
\toprule[1.25pt]
\multicolumn{3}{c}{\textbf{Yahoo}} \\
\toprule[1.25pt]
\textbf{Dataset/Case}~    & \textbf{Ill-formed}              & \textbf{Well-formed}                    \\ 
\midrule[1pt]
          & \texttt{what's the \underline{differncee}} & \textbf{Seq2Seq}:$\,\,$ \texttt{what's the  \underline{difference} between human beings and cancer cancer?}  \\ 
\cline{3-3}
 Yahoo/Case 1     & \texttt{between climate \underline{chnage}}  &  \textbf{TOQR}:$\,\,$ \texttt{  \underline{differncee} between climate  \underline{chnage} and global  \underline{wraming}}          \\ 

Wrong word  & \texttt{and global \underline{wraming}?}             & $\quad \quad \quad$ \texttt{ oxygen weather  \underline{warming}}                        \\ 
\cline{3-3}

 & ~              & \textbf{AQA}: $\,\,$\texttt{what's the  \underline{difference} between global and global  \underline{warming} war?}                        \\ 
\cline{3-3}
            &                         &   \textbf{QREFINE}: $\,\,$\texttt{what's the  \underline{difference} between climate  \underline{change} and global  \underline{warming}?}                          \\
\midrule[1pt]
            & \texttt{how safe to meet is }   & \textbf{Seq2Seq}: $\,\,$ \texttt{how to make it is an adult professional?}                     \\ 
\cline{3-3}
Yahoo/Case 2      &  \texttt{it an american?}                  &\textbf{TOQR}:$\,\,$ \texttt{how safe meet an american again powerful secretary}                       \\ \cline{3-3}
Wrong order &                         & \textbf{AQA}: $\,\,$\texttt{an animal how safe is it to meet an american?}                      \\ 
\cline{3-3}
            &                         &    \textbf{QREFINE}: $\,\,$\texttt{how safe is it to meet an american?}                     \\ 
\midrule[1pt]
            &     \texttt{who would able help }              & \textbf{Seq2Seq}:$\,\,$ \texttt{what is empirical and how is it used mechanics?}                     \\ 
\cline{3-3}
Yahoo/Case 3      &  \texttt{what is string theory}                 &\textbf{TOQR}:$\,\,$ \texttt{able help string theory and how is it used}                      \\ 
Noisy Background &    \texttt{and how is it used?}                    & $\quad \quad \quad$ \texttt{theory physics used model}                         \\ 
\cline{3-3}

 &                        & \textbf{AQA}: $\,\,$\texttt{what is turbo motion and how it is used?}                         \\ 
\cline{3-3}
            &                         &    \textbf{QREFINE}: $\,\,$\texttt{what is string theory and how is it used?}                      \\
\bottomrule[1.25pt]
\multicolumn{3}{c}{}\\
\toprule[1.25pt]
\multicolumn{3}{c}{\textbf{CSU}} \\
\toprule[1.25pt] 
\textbf{Dataset/Case}~    & \textbf{Ill-formed}              & \textbf{Well-formed}                    \\ 
\midrule[1pt]
                 & {\small \begin{CJK}{UTF8}{gbsn}我想问一下为什么\end{CJK}}                            & \textbf{Seq2Seq}: $\,\,$ {\small \begin{CJK}{UTF8}{gbsn}如何使用\underline{优惠券}\end{CJK}} \quad \quad \texttt{How to use the \underline{coupon}}                           \\ \cline{3-3} 
CSU/Case 1       & {\small \begin{CJK}{UTF8}{gbsn}\underline{优惠卷}用不了\end{CJK}} & 

\textbf{TOQR}: $\,\,$
                 {\small \begin{CJK}{UTF8}{gbsn} \underline{优惠卷}  \underline{优惠券} 使用规则 有效期 特价商品\end{CJK}}   \\ 
 Wrong word                  & \texttt{I want to ask why}  & 
             $\quad \quad \quad$    \texttt{ \underline{coubon}  \underline{coupon} service regulations validity bargain goods} 

\\ \cline{3-3}

      &       \texttt{\underline{coubon} cannot be used}                                                                                  & \textbf{AQA}: $\,\,$ {\small \begin{CJK}{UTF8}{gbsn} 如何获得\underline{优惠券}\end{CJK}} \quad \quad \texttt{how can I get \underline{coupon}}                           \\ \cline{3-3} 
                 &               & 
                 
                 \textbf{QREFINE}: {\small \begin{CJK}{UTF8}{gbsn}\underline{优惠券}不能使用怎么办\end{CJK}} \quad \quad \texttt{What should I do if the \underline{coupon} cannot be used}                                                
                
                 \\
\midrule[1pt]
                 & {\small \begin{CJK}{UTF8}{gbsn} 我怎么办手机进水了 \end{CJK}}                         & \textbf{Seq2Seq}:     {\small \begin{CJK}{UTF8}{gbsn}手机丢失怎么找回\end{CJK}} \quad \quad \texttt{how do I find the lost phone}                                      \\
CSU/Case 2       & \texttt{I do what phone has}                                                                                       & \textbf{TOQR}: $\,\,$  {\small \begin{CJK}{UTF8}{gbsn} 手机进水 远离水源 吸水 \end{CJK}}    \quad \quad \texttt{water damage avoid water absorb water}   \\ \cline{3-3}
               
Wrong order      & \texttt{water damage}                                                                                              & \textbf{AQA}: $\,\,$   {\small \begin{CJK}{UTF8}{gbsn} 手机进水了 \end{CJK}} \quad \quad \texttt{my phone has water damage}                                                                                   \\ \cline{3-3} 
                 &                                                                                                                                     &     \textbf{QREFINE}: $\,\,$   {\small \begin{CJK}{UTF8}{gbsn}手机进水怎么办\end{CJK}} \quad \quad \texttt{What should I do if the phone has water damage}                    \\ 
\midrule[1pt]
                 & {\small \begin{CJK}{UTF8}{gbsn} 忘记密码多次 \end{CJK}}                            & \textbf{Seq2Seq}:   $\,\,$      {\small \begin{CJK}{UTF8}{gbsn} 无法登陆怎么办 \end{CJK}} \quad \quad \texttt{What should I do when cannot log in}                                                              \\ \cline{3-3} 
CSU/Case 3       & {\small \begin{CJK}{UTF8}{gbsn}登录现在没法登陆 \end{CJK}}                           & \textbf{TOQR}: {\small \begin{CJK}{UTF8}{gbsn}  忘记密码 没法登陆 账号冻结 违规 错误密码 \end{CJK}}  \\ 
Noisy Background& \texttt{forgot the password and }    & $\quad \quad \quad$
\texttt{forget password cannot login block account number wrong password}  \\ \cline{3-3} 
 &           \texttt{repeatedly log in right}                                                           & \textbf{AQA}: $\,\,$  {\small \begin{CJK}{UTF8}{gbsn} 忘记密码怎么办 \end{CJK}} \quad \quad \texttt{What should I do when I forget password}                                                                    \\ \cline{3-3} 
                 &       \texttt{now cannot log in}                                                                     & \textbf{QREFINE}:     {\small \begin{CJK}{UTF8}{gbsn} 登陆密码忘记了怎么办 \end{CJK}} \quad \quad  \texttt{What should I do when login password forgot}                                                        \\ 
\bottomrule[1.25pt]
\end{tabular}}
\caption{Cases study of Generated Results on Yahoo dataset and Commercial Customer Service Userlog dataset. Typos and substitutions are shown in underscore.}
\label{yahoo-case}
\end{table*}

\subsubsection{Case Study}
To demonstrate the readability and effectiveness of the proposed method, Table \ref{yahoo-case} shows examples of generated outputs on Yahoo and CSU dataset. As we can see, {\ModelName} has the ability to correct the wrong words (shown in Case 1), change the question word order for better readability (shown in Case 2) and remove the unnecessary background in the original question (shown in Case 3). For example, ``differncee'' can be correct as ``difference''. And the refined question by our model is readable. But the question generated by other baselines cannot well express the original meaning in the ill-formed question or be misleading and also have problems like repeatedly the useless words (e.g., Seq2Seq in Yahoo/Case 1), no natural language sentence (e.g., TOQR in Yahoo/Case 1) and express the different meaning with the ill-formed question (e.g., AQA in Yahoo/Case 1). Therefore, the question generated by {\ModelName} is more readable than other alternatives and is able to keep the original user intention of asking the question.

\subsection{Answer Retrieval}
To validate the effectiveness of question refinement in helping retrieve answers in existing QA systems, we use PyLucene \footnote{\url{http://http://lucene.apache.org/pylucene/}} for retrieving the answer to the search question.\\
\noindent\textbf{Hits@K}:
The top K relevant answers retrieved by the search question using PyLucene. If the gold answer is inside of the top K retrieved answers, then the Hits@K of this search question equals 1, otherwise, it equals 0. The whole Hits@K of questions is the average development set question's Hits@K.\\
\noindent \textbf{Results}
In Table \ref{QA_retrieval}, we can see the all refined question generated can be better than the Ill-formed question, which shows that the refinement process is needed. TOQR, which aims to maximize the retrieval results achieves the good performance compared with other methods. However, our model {\ModelName} achieves very better performance comparing with TOQR in most case and the question refined by our model has better readability. The Hits@K score retrieved by {\ModelName}-word is higher than Seq2Seq, which indicates that by improving the readability of question, the retrieval ability also be improved. As {\ModelName}-sen performs better than Seq2Seq, it shows that the reward considering over-all question structure for a better correlation with refined question to its answer is important.
This result shows the superiority of {\ModelName} in greatly improving QA utility via explicitly refining the question for enhancing its readability and retrieve ability for both computer and human. \\

\subsection{Ablation Study}
In order to find out which part of the model improves the automatic evaluation performance, we do the ablation study. \textbf{S2S+W} is seq2seq model using word-level embedding. \textbf{S2S+W\&C} is seq2seq model using word-level and char-level embedding. \textbf{S2S+W\&C\&B} is seq2seq model using word-level, char-level, and BERT embedding. \textbf{QR-word} is seq2seq model considering three embeddings and using word reward to train the RL model. \textbf{QR-ans} is seq2seq model considering three embeddings and using answer coherence reward to train the model. \textbf{QR-RF} considers multi-grain word embedding and both word reward and sentence reward but uses REINFORCE method. \textbf{QR-PPO} is our model. 

\begin{figure}[t!]
\centering
  \subfigure[Yahoo]{\label{Yab}
    \begin{minipage}[l]{.475\columnwidth}
      \centering
      \includegraphics[width=\textwidth]{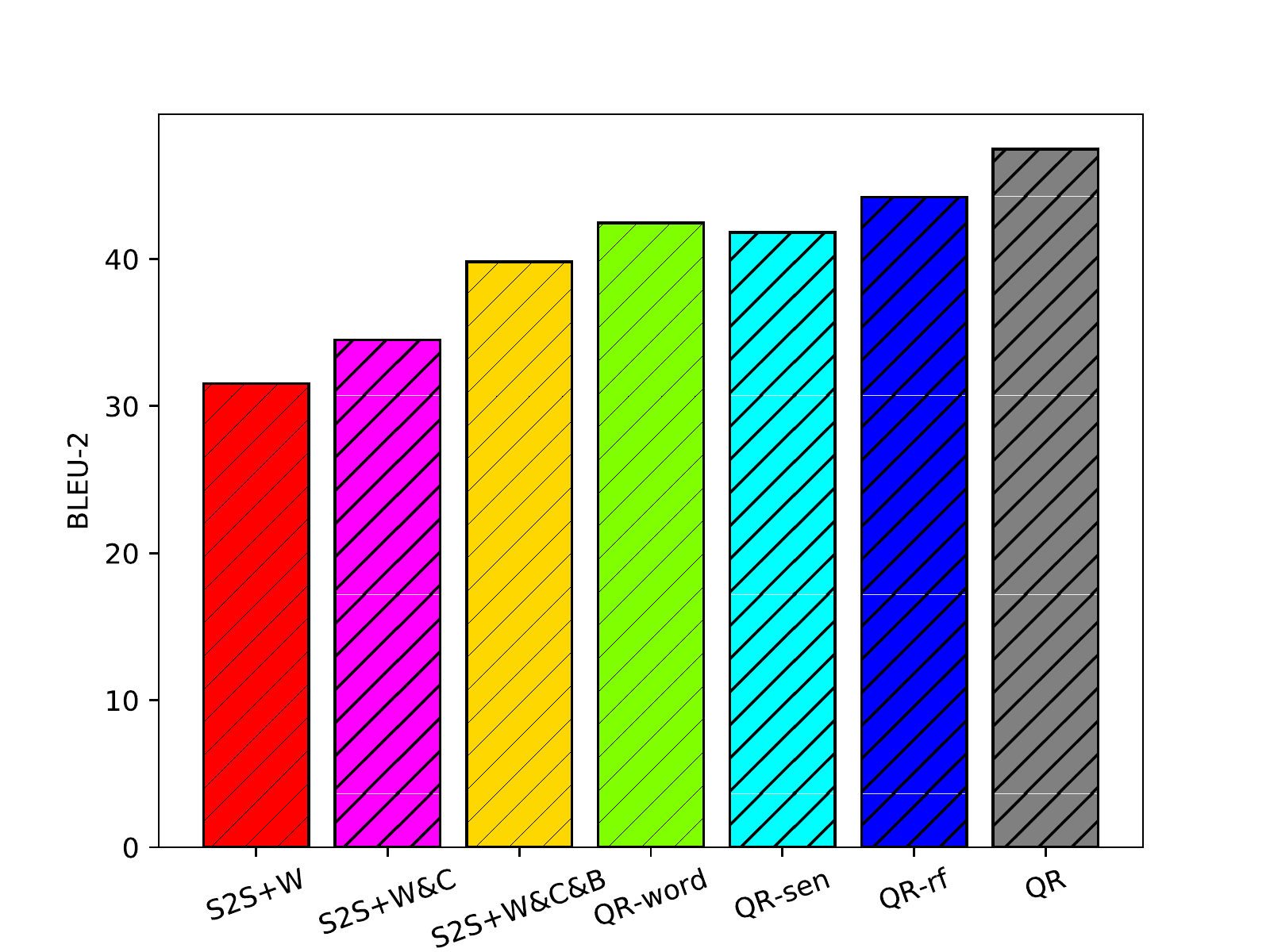}
    \end{minipage}
  }
\subfigure[CSU]{ \label{Cab}
    \begin{minipage}[l]{.475\columnwidth}
      \centering
      \includegraphics[width=\textwidth]{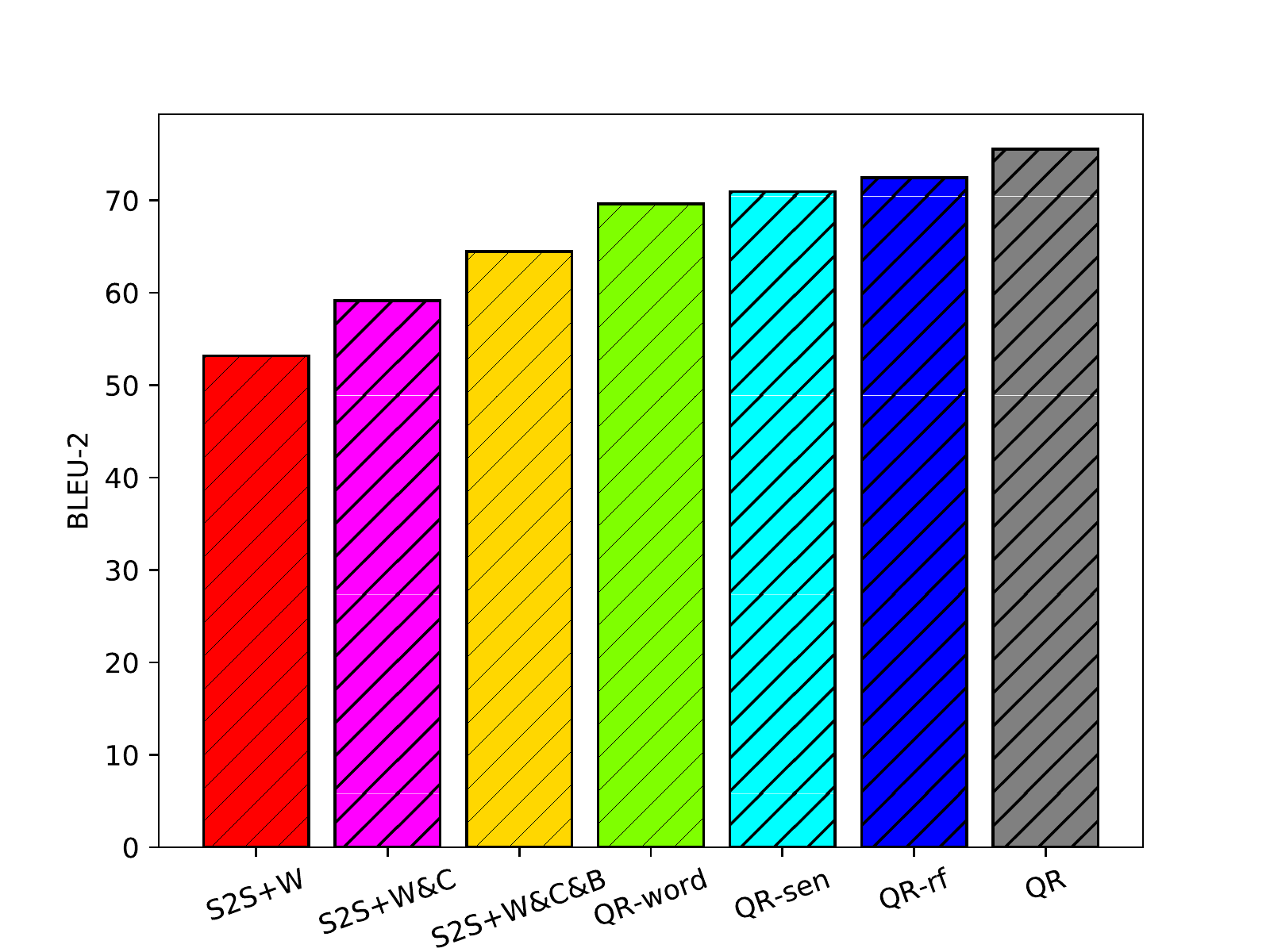}
    \end{minipage}
  }
  \caption{The ablation study on Yahoo and CSU dataset}
\end{figure}

From Fig. \ref{Yab} and \ref{Cab}, we can see that using multi-grain word embedding can help the model better to correct the ill-formed question than just using single word embedding. And using PPO reinforcement learning with the word-level and sentence-level reward can improve the model to learn a stable policy that can generate the appropriate well-formed question. 

\begin{table}[t!]
\centering
\resizebox{0.75\linewidth}{!}{
\begin{tabular}{l|cccc}
\toprule[1.25pt]
\textbf{Yahoo}      & \textbf{Hits@1} & \textbf{Hits@3} & \textbf{Hits@5}  & \textbf{Hits@10} \\
\hline
Ill-formed  & 3.35 & 4.89 & 8.68 & 14.98 \\
Seq2Seq & 3.60 & 4.50 &8.99	&16.97 	 \\
PARA-NMT  & 6.47 & 7.95 & 12.01 & 17.21 \\
AQA & 6.48 & 8.12&	13.06 & 18.78 \\ 
TOQR & 7.27 & 10.25	& 22.83 &\textbf{32.43}	\\
\hline
QR-word & 7.11	&9.34	&20.79&	29.88 \\
QR-ans & 7.16	&9.67	&22.45 & 30.89 \\ 
{\ModelName} &\textbf{8.09} &\textbf{10.76}&\textbf{23.95}  & 32.23	 \\
\bottomrule[1.25pt]
\multicolumn{4}{c}{} \\
\toprule[1.25pt]
\textbf{CSU}     & \textbf{Hits@1}   & \textbf{Hits@3} & \textbf{Hits@5} & \textbf{Hits@10} \\
\hline
Ill-formed & 10.34 & 18.72 & 26.78 & 39.78  \\
Seq2Seq &11.84 & 19.83	&27.25	&40.12		 \\
PARA-NMT & 15.59 & 21.09 & 30.95 & 40.33  \\
AQA & 16.89 &20.00	& 31.81 &40.67		\\
TOQR & 20.41 & 27.98	& 34.59	&48.78		 \\
\hline
QR-word	&21.23 	&25.98	&33.04 &46.19	 \\
QR-ans	&19.64 	&26.11	&34.67 &47.76	 \\
{\ModelName} &\textbf{22.10} &\textbf{28.69}	&\textbf{35.98} &\textbf{49.16}		 \\
\bottomrule[1.25pt]
\end{tabular}}
\caption{Answering Retrieval Result on Yahoo and CSU dataset.}\label{QA_retrieval}
\end{table}

\subsection{Learning Curves Analysis} \label{sub:lc}
\begin{figure}[t]
\centering
  \subfigure[Yahoo]{\label{Ylc}
    \begin{minipage}[l]{.475\columnwidth}
      \centering
      \includegraphics[width=\textwidth]{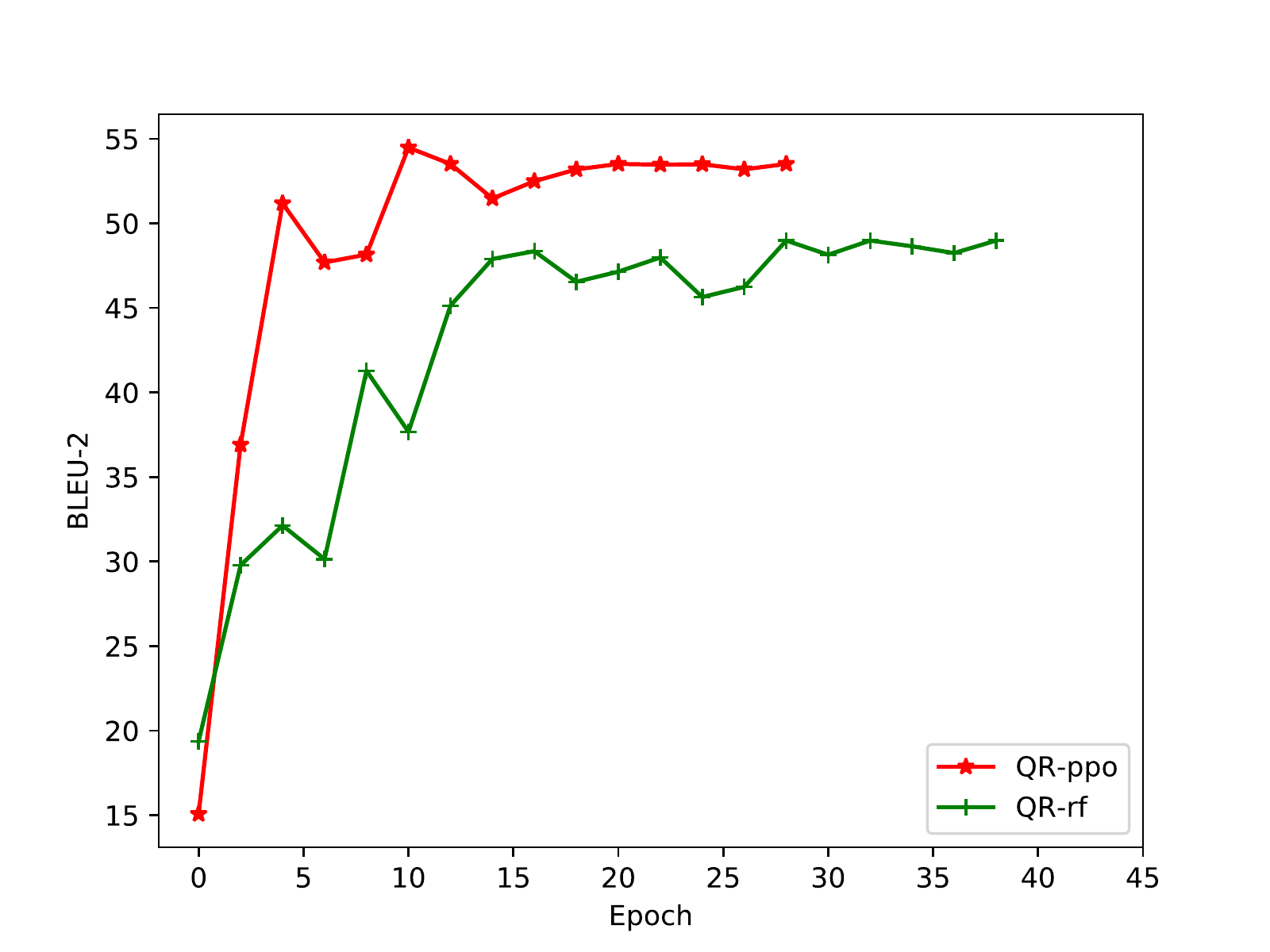}
    \end{minipage}
  }
\subfigure[CSU]{ \label{Clc}
    \begin{minipage}[l]{.475\columnwidth}
      \centering
      \includegraphics[width=\textwidth]{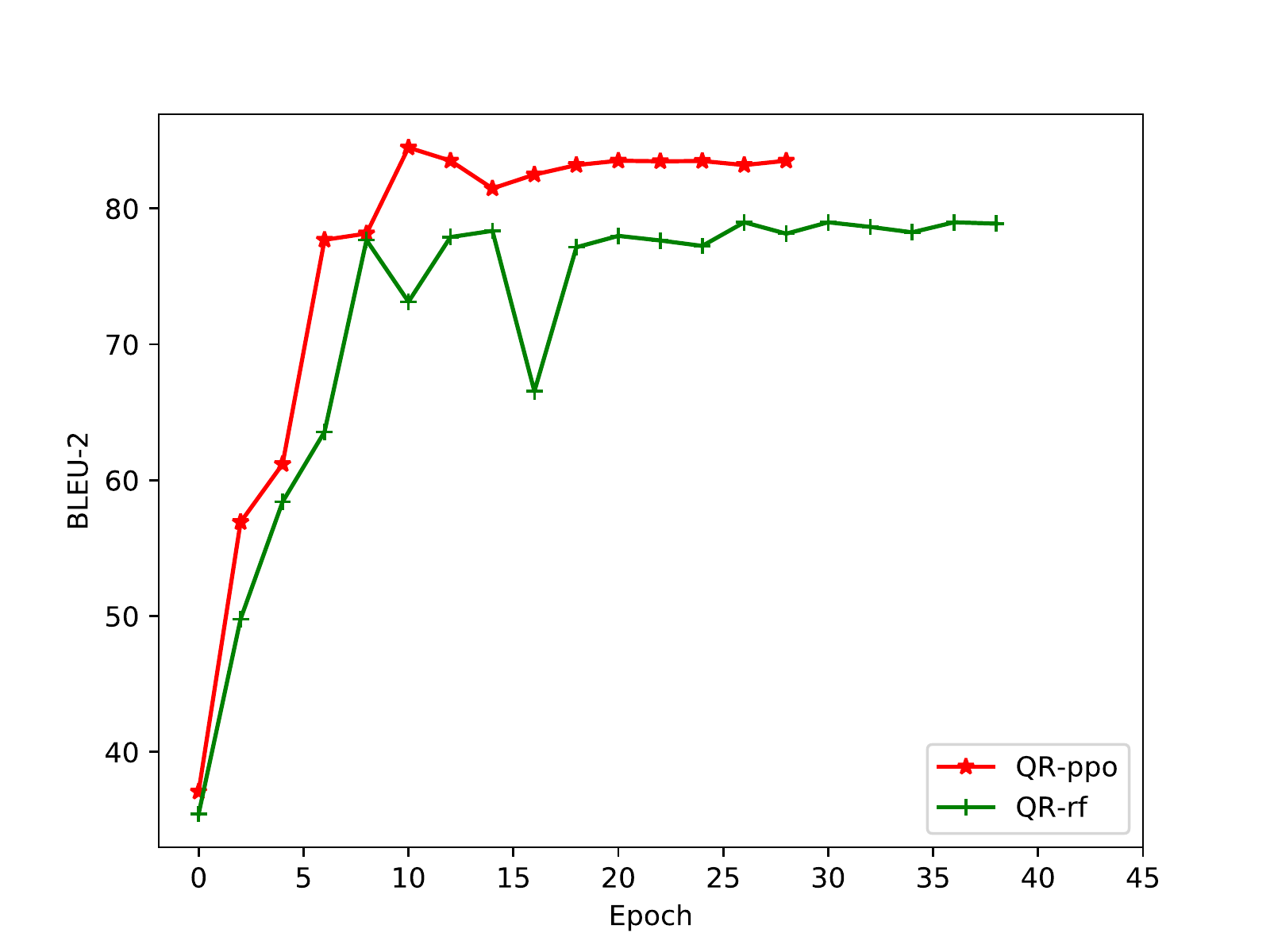}
    \end{minipage}
  }
  \caption{The learning curve analysis on Yahoo and CSU dataset}
\end{figure}

The BLEU-2 scores and learning curves of different optimization algorithms are presented in Figure \ref{Ylc} and \ref{Clc}. From the testing results in Table \ref{bleu_score}, we can see that the two optimization methods have comparable performance, but PPO achieves a slightly higher BLEU-2 score than REINFORCE. Moreover, we find out that the training progress of PPO is more stable and converge earlier than policy gradient. This shows that PPO methods can improve the high variance problem of using REINFORCE, and can help the learning converge quickly.

\section{Related Work}\label{sec:related}
\subsection{Generative Text Refinement}
The generative ability of deep neural networks leads to the prevalence of Seq2Seq models for reformulation task. These methods typically accomplish the reformulation by training a recurrent neural network such as an LSTM network to predict the next word in a sentence sequence. \cite{nogueira2017task} uses the reinforcement learning to reformulate the query to maximize the number of relevant documents retrieved. Our work differs with their method in that we generate natural language sequences rather than terms; thus their reformulated query doesn't contain the propriety of readability and understandability. \cite{zhang2017sentence} proposed the sentence simplification task, which aims to make sentences easier to read and understand. Similarly, \cite{xie2016neural} operates on the character level to flexibly handle orthographic errors in spelling, capitalization, and punctuation. Although their frameworks can reform the sentences to be more readable, their objective does not include the capability of refining the question to get answers easier. Active QA (AQA) \cite{buck2018ask} uses an agent as a mediator between the user and a black box QA system, e.g. BiDAF, to generate questions that elicit the best possible answer. Since the pretrained fixed environment, BiDAF, is not updating with the model, feedback on the quality of the question reformulations could be quite noisy which presents a challenge for training. Moreover, BiDAF works on the reading comprehension, the answer is the paraphrase in the context, therefore there is a great change that this model always generates the same answer, which brings another challenge for training. 
\subsection{Reinforcement Learning for QA}
Due to the high dimensional action space for text generation and high diversity of the required generation result, policy gradient methods are more appropriate in the text generation than value-based methods like Q-learning \cite{mnih2013playing}. By using policy gradient method, the limitation of cross-entropy loss that inherently comes with word-level optimization is alleviated and allowing sequence-level reward functions, like BLEU, to be used \cite{ranzato2015sequence}. \cite{bahdanau2016actor} extends this line of work using actor-critic training. Uses of policy gradient for QA include \cite{liang2016neural}, who train a semantic parser to query a knowledge base, and \cite{seo2016query} who propose query reduction networks that transform a query to answer questions that involve multi-hop common sense reasoning. Li et al. \cite{li2017paraphrase} use RL and SL to learn the paraphrase of the sentence. The on-policy method like REINFORCE suffers the high variance and slow to converge. The off-policy method like TRPO and PPO \cite{schulman2015trust, schulman2017proximal} recently applied on the game like Atari. They can deal with the problems by regularizing the gradient of policy. Tuan et al \cite{tuan2018proximal} apply the off-policy gradient method to the sequence generation task and shows that PPO surpass policy gradient on stability and performance.

\section{Conclusion and Future Work}
Question refinement aims to refine ill-formed questions, which typically includes various types of subtasks such as spelling error correction, background removal and word order refinement. Instead of tackle these subtasks separately, we develop a unified model, based on Seq2Seq, to handle this task in a data-driven way. We improve the question representation by incorporating character embedding and contextual word embedding such as BERT. To make the refinement process more controllable, we combine Seq2Seq model with deep reinforcement learning. We define a sequence generator by optimizing for a combination of imposed reward functions. The experimental results show that our method can not only produce more readable question but also significantly improves the retrieval ability of question for downstream QA system.

Question refinement is a challenging task and there are several directions to improve. One direction is to develop the advanced method, such as creating different awards that are more suitable to deal with the three subtasks. Besides, In our setting, the ill-formed and well-formed questions still need to be paired. In most of realistic cases, we only have a pool of well-formed. We seek to use inverse reinforcement learning \cite{wang2018no} to learn the intrinsic representation of the well-formed question. Therefore, given an ill-formed question, the model can refine it to the well-formed. Finally, it is also interesting to make use of the result of question refinement to improve other related tasks such as question understanding and question answering recommendation.

\section*{Acknowledgements}
This work is supported in part by NSF under grants III-1526499, III-1763325, III-1909323, SaTC-1930941, and CNS-1626432. 


\end{document}